\documentclass[prl,twocolumn,nofootinbib]{revtex4}
\usepackage{graphicx}
\usepackage{dcolumn}
\usepackage{amssymb}
\usepackage{bm}
\usepackage{enumerate}
\def\be{\begin{enumerate}}
\def\ee{\end{enumerate}}
\def\beq{\begin{equation}}
\def\eeq{\end{equation}}
\def\bea{\begin{eqnarray}}
\def\eea{\end{eqnarray}}
\def\a{\alpha}

\def\g{\gamma}
\def\d{\delta}

\def\3halfs{\textstyle{\frac{3}{2}}} \def\em{\it}
\newcommand{\sla}[1]%
        {{\raise.15ex\hbox{$/$}\kern-.57em #1}}
\newcommand{\Sla}[1]%
        {{\raise.15ex\hbox{$/$}\kern-.75em #1}}

\def\ben{\begin{enumerate}}
\def\een{\end{enumerate}}
\def\bitem{\begin{itemize}}
\def\eitem{\end{itemize}}

\def\a{\alpha}

\def\g{\gamma}

\begin{document}
\def\Universita{Universit\`a}
\title{New limits on Planck scale Lorentz violation in QED}
\author{T. Jacobson}
\affiliation{Department of Physics University of Maryland, College Park, MD
20742-4111, USA}
\author{S. Liberati}
\altaffiliation[Present address: ]{SISSA and INFN, Trieste, ITALY}
\affiliation{Department of Physics University of Maryland, College Park, MD
20742-4111, USA}
\author{D. Mattingly}
\altaffiliation[Present address: ]{Dept. of
Physics, U.C.~Davis, CA, USA}
\affiliation{Department of Physics University of Maryland, College Park, MD
20742-4111, USA}
\author{F.W. Stecker}
\affiliation{LHEA, NASA Goddard Space Flight Center, Greenbelt, MD 20771, USA}


\begin{abstract}
Constraints on possible Lorentz symmetry violation (LV) of order
$E/M_{\rm Planck}$ for electrons and photons in the framework of
effective field theory (EFT) are discussed.  Using (i) the report of polarized
MeV emission from GRB021206 and (ii) the absence of vacuum
\v{C}erenkov radiation from synchrotron electrons in the Crab
nebula, we improve previous bounds by $10^{-10}$ and $10^{-2}$
respectively.  We also show that the LV parameters for positrons
and electrons are different, discuss electron helicity decay, and
investigate investigate how prior constraints are modified by the
relations between LV parameters implied by EFT.

\end{abstract}

\maketitle

The past few years have witnessed a rapid development of
powerful constraints on some types of Lorentz symmetry
violation (LV) that have been suggested by quantum
gravity scenarios.   While no current suggestion of LV
is firm enough to be considered a prediction, there is
nevertheless great interest in the possibility of LV
induced by Planck scale physics since it offers the hope
of an observational window into quantum gravity.  To date
no LV phenomena have been observed  (although
the ultra high energy cosmic ray events detected by the
Akeno Giant Air Shower Array (AGASA), could possibly
turn out to be harbingers of LV physics~\cite{CG}).  The
absence of LV provides important constraints on viable
quantum gravity theories.  Moreover, these constraints
are interesting in their own right as they extend the
domain where relativity has been tested far beyond its
previous frontiers.

The primary purpose of this paper is to further strengthen the
bounds on LV of order $E/M_P$ for photons and electrons, where
$M_P=(\hbar c^5/G)^{1/2}=1.22\times10^{19}$ GeV is the Planck
energy, the presumed energy scale of quantum gravity.  We use the
reported observation~\cite{GRBpol} of polarized gamma rays from 
the gamma ray burst GRB021206 to improve the birefringence constraint by ten
orders of magnitude. [The results of~\cite{GRBpol} have
been challenged~\cite{challenge} and defended~\cite{CBResponse}.
If the polarization turns out to be weaker, then the birefringence
constraint from GRB021206 is weakened (or eliminated).] 
By consideration of the vacuum
\v{C}erenkov process for the electrons producing the highest
frequency synchrotron radiation from the Crab nebula we improve
on the old birefringence constraint by two orders of magnitude.

A secondary purpose is to revisit previous constraints in
light of the effective field theory (EFT) analysis
of~\cite{MP}, some of which are strengthened and some
weakened or limited in applicability.  We show that EFT
implies that the LV parameters for positrons are opposite
(in two senses) compared to electrons, and we discuss a
new LV process of ``helicity decay", in which an
electron of one helicity decays to a state with the
opposite helicity. Finally we pull together the strongest
constraints to date and present them in a logarithmic
plot that allows their nature and relative strength to be
easily compared to previous work.

We adopt the framework of effective field theory as
developed e.g.~in~\cite{CK,CG,MP}, focusing on the
electron-photon sector since this involves no other
particles and there are many observations allowing a
number of independent constraints to be combined.  We
assume rotational symmetry is preserved in a preferred
frame, which is taken to coincide with that of the cosmic
microwave background radiation, and consider only LV
suppressed by one power of the ratio $E/M_{\rm Planck}$,
which arises from mass dimension five operators in the
Lagrangian.
(We thus assume that lower mass dimension
LV operators are suppressed by a symmetry or other
mechanism, otherwise
they would be expected to dominate~\cite{MP,Perez}.)

Under these assumptions the most general
photon and electron dispersion relations are~\cite{MP}
\bea
E^2 &=& p^2 \pm \xi\,  p^3/M \qquad~~~~~~~~~~~~{\rm photons}
\label{eq:disp-ph}\\
E^2 &=& m^2 + p^2 + \eta_{R,L}\, p^3/M \qquad {\rm electrons}
\label{electrons}
\eea
where $\xi$, $\eta_R$, and $\eta_L$ are independent
dimensionless parameters, and
$M=10^{19}$ GeV is factored out rather than the
Planck mass $M_P=1.22 M$ for computational
convenience. We adopt units with $\hbar=1$ and the low
energy speed of light $c=1$.
The sign in the photon dispersion relation
(\ref{eq:disp-ph}) corresponds to the helicity
(i.e. right or left circular polarization), while the
labels $R$ and $L$ in the electron dispersion relation
(\ref{electrons}) apply for positive and negative
electron helicity respectively (see below for more
details).  The bound $|\eta_L -\eta_R|\leq 4$~\cite{MP}
is provided by measurements of spin-polarized torsion
pendulum frequency~\cite{eotwash}.

{\em New birefringence constraint}.--- The dispersion relation
(\ref{eq:disp-ph}) implies that electromagnetic waves of
opposite helicity have different phase velocities, which
leads to a  rotation of linear polarization direction
through the angle
\beq \theta(t)=\left[\omega_+(k)-\omega_-(k)\right]t/2=\xi k^2 t/2M
\label{rotation}
\eeq
for a plane wave with wave-vector $k$. Observations of
polarized radiation from distant sources can hence be
used to place an upper bound on $\xi$.

The best previous bound, $|\xi|\lesssim 2\times 10^{-4}$,
was obtained by Gleiser and Kozameh~\cite{GK}, using the observed 10\%
polarization
of ultraviolet light
from a distant galaxy.
(See also \cite{CFJ,KM} for similar birefringence bounds
in the context of different types of Lorentz symmetry
breaking.)


Recently the prompt emission from the gamma ray burst 
GRB021206 was observed using the RHESSI detector~\cite{rhessi}. 
A linear polarization of 80\% $\pm$ 20\% was reported~\cite{GRBpol}.
[This claim has been challenged~\cite{challenge} and 
defended~\cite{CBResponse}.]
During the five seconds of emission the intensity varied 
strongly on a timescale of small fractions of a second 
consistently across the spectral window 0.15-2 MeV. 
The data~\cite{rhessidata} indicate a major contribution to the flux 
comes from photons significantly distributed over 
at least the energy range 0.1-0.5 MeV.

The constraint arises from the fact that
if the angle of polarization rotation (\ref{rotation})
were to differ by more than $\pi/2$ over the
range 0.1-0.3 MeV (and hence by more than
$3\pi/2$ over the range 0.1-0.5 MeV),
the instantaneous polarization
at the detector would fluctuate sufficiently
for the net
polarization of the signal to be suppressed well
below the observed value. (A stronger constraint could
clearly be obtained by taking into account more precisely
the spectral characteristics of the signal and detector.)
The difference in rotation angles for wave-vectors $k_1$
and $k_2$ is
\beq
 \Delta\theta=\xi (k_2^2-k_1^2) d/2M,
  \label{diffrotation}
\eeq
where we have replaced the time $t$ by the distance $d$
from the source to the detector (divided by the speed of
light).

While the distance to GRB021206 is unknown, it is well
known that most cosmological bursts have redshifts in the
range 1-2 corresponding to distances of greater than a
Gpc. Using the distance distribution derived in
Ref.~\cite{DLRG} we conservatively take the minimum
distance to this burst as 0.5 Gpc, corresponding to a
redshift of $\sim 0.1$.
This then yields the constraint
\beq
|\xi|<5.0\times10^{-15}/d_{0.5}.
\eeq
where $d_{0.5}$ is the distance to the burst in units of 0.5 Gpc.

{\em New \v{C}erenkov-Synchrotron constraint}.--- In a
region of the LV parameter space there is an energy
threshold for a free electron to emit a photon in a
process called vacuum \v{C}erenkov radiation. The
threshold can occur either with emission of a soft photon
or a hard photon depending on the
parameters~\cite{Long,KonMaj}.  In the soft photon case
the threshold is
$E_{\rm th} = (m^2M/2\eta)^{1/3}\approx 11~{\rm TeV}/\eta^{1/3}$, from
which it follows that the strength of the constraint on
$\eta$ scales as the inverse cube of the electron energy, and
that energies of order 10 TeV for the electron are
required in order to put constraints of order unity on
the LV parameters~\cite{Long,KonMaj}.

Electrons of energy up to 50 TeV are inferred via the
observation of 50 TeV gamma rays from the Crab nebula
which are explained by inverse Compton (IC) scattering.  Since
the \v{C}erenkov rate is orders of magnitude higher than
the IC scattering rate, the \v{C}erenkov
process must not occur for these electrons~\cite{CG,Long}.
This yields a constraint on $\eta$ of order $(10~{\rm
  TeV}/50~{\rm TeV})^3\sim 10^{-2}$.  Neither photon
helicity should be emitted, so the absolute value $|\xi|$
is bounded, which strengthens the  IC \v{C}erenkov
constraint. On the other hand, it could be that only one
electron helicity produces the IC photons
and the other loses energy by vacuum \v{C}erenkov
radiation. Hence we can infer only that at least one of
$\eta_R$ and $\eta_L$ satisfies the bound.

A complementary constraint was derived in~\cite{Crab} by making use
of the very high energy electrons that produce the
highest frequency synchrotron radiation in the Crab
nebula.  For negative values of $\eta$ the electron has a
maximal group velocity less than the speed of light,
hence there is a maximal synchrotron frequency that can
be produced regardless of the electron
energy~\cite{Crab}. Observations of the Crab nebula
reveal synchrotron radiation at least out to 100 MeV
(requiring electrons of energy 1500 TeV in the Lorentz
invariant case), which implies that at least one of the
two parameters $\eta_{R,L}$ must be greater than
$-7\times10^{-8}$ (this constraint is independent of the
value of $\xi$).  We cannot constrain both $\eta$
parameters in this way since it could be that all the
Crab synchrotron radiation is produced by electrons of
one helicity.
Hence for the rest of this discussion let
$\eta$ stand for whichever of the two $\eta$'s satisfies
the synchrotron constraint.

This must be the same $\eta$ as satisfies the
IC \v{C}erenkov constraint discussed
above, since otherwise the
energy of these synchrotron electrons
would be below 50 TeV rather than
the Lorentz invariant value of 1500 TeV.
The Crab spectrum is well accounted for
with a single population of electrons responsible
for both the synchrotron radiation and the
IC $\g$-rays. If there were enough extra
electrons to produce the observed synchrotron flux
with thirty times less energy per electron,
then the electrons of the other helicity
which would be producing the
IC $\g$-rays would be too numerous.

We now use the existence of these synchrotron producing electrons
to improve on the vacuum \v{C}erenkov constraint. For a given
$\eta>0$, some definite electron energy $E_{\rm synch}(\eta)$ must
be present to produce the observed synchrotron radiation. (This is
higher for negative $\eta$ and lower for positive $\eta$ than the
Lorentz invariant value~\cite{Crab}.) Values of $|\xi|$ for which
the vacuum \v{C}erenkov threshold is lower than $E_{\rm
synch}(\eta)$  for either photon helicity can therefore be
excluded. (This is always a hard photon threshold, since the soft
photon threshold occurs when the electron group velocity reaches
the low energy speed of light, whereas the velocity required to
produce any finite synchrotron frequency is smaller than this.)
For negative $\eta$, the \v{C}erenkov process occurs only when
$\xi<\eta$~\cite{Long,KonMaj}, so the excluded parameters lie in
the region $|\xi|>-\eta$.

 {\em Implications of EFT for prior constraints}.---
 
\begin{figure}[t]
{\includegraphics[scale=0.85]{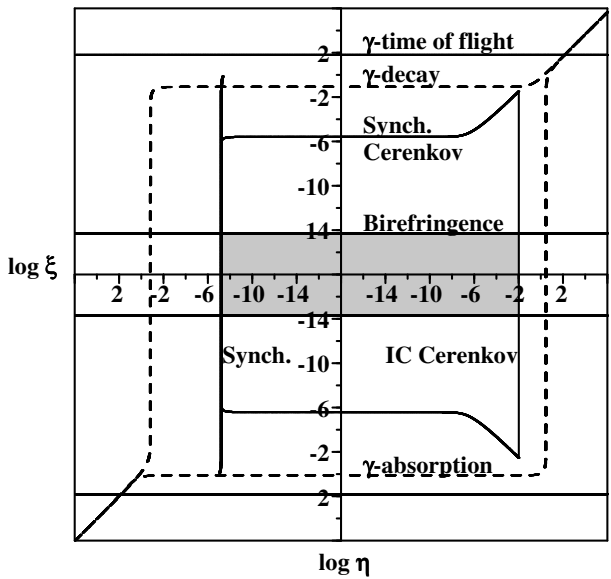}}
\caption{Constraints on the photon ($\xi$) and electron ($\eta$)
LV parameters.  The birefringence constraint uses the observed
polarization of MeV photons from GRB021206.  The synchrotron and
IC Cerenkov constraints use the observation of 0.1 GeV synchrotron
and 50 TeV inverse Compton radiation from the Crab nebula,
respectively. For the origin of other constraints see text. For
negative parameters the negative of the logarithm of the absolute value is
plotted, and a region of width $10^{-18}$ is excised around each
axis. The synchrotron and \v{C}erenkov constraints are known to
apply only for at least one $\eta_{R,L}$. The IC and synchrotron
\v{C}erenkov lines are truncated where they cross. Prior photon
decay and absorption constraints are shown in dashed lines since
they do not account for the EFT relations between the LV
parameters.}
\label{fig:all}
\end{figure}

{Photon time of flight.}---The Lorentz violating dispersion relation
(\ref{eq:disp-ph}) implies that the group velocity of photons,
$v_{g}=1\pm\xi p/M$, is energy dependent. This leads to an energy
dependent dispersion in the arrival time at Earth for photons
originating in a distant event~\cite{Pav,ACNature}, which was
previously exploited for
constraints~\cite{Schaefer,Biller,Kaaret}. The dispersion of the
two polarizations is larger since the difference in group velocity
is then $2|\xi|p/M$ rather than $\xi(p_2-p_1)/M$, but the time of
flight constraint remains many orders of magnitude weaker than the
birefringence one from polarization rotation. In
Fig.~\ref{fig:all} we use the EFT improvement of the constraint
of~\cite{Biller} which yields $|\xi|<63$.

 {Photon decay and photon absorption.}---The constraints from photon decay
$\g\rightarrow e^+e^-$ and absorption $\g\g\rightarrow e^+e^-$
must be reanalyzed to take into account the different dispersion
for the two photon helicities, and the different parameters for
the two electron helicities, but there is a further complication:
both these processes involve positrons in addition to electrons.
Previous constraint derivations have assumed that these have the
same dispersion, but that need not be the case~\cite{Lehnert}.
We show below for the $O(E/M)$ corrections that it is indeed not
so. Taking into account the above factors could not significantly
improve the strength of the constraints (which is mainly
determined by the energy of the photons). We indicate here only
what the helicity dependence of the photon dispersion implies,
neglecting the important role of differing parameters for
electrons and positrons and their helicity states.

The strongest limit on photon decay came from the highest energy
photons known to propagate, which at the moment are the 50 TeV
photons observed from the Crab nebula~\cite{Long,KonMaj}.  Since
their helicity is not measured, only those values of $|\xi|$ for
which {\it
  both} helicities decay could be ruled out.
The photon absorption constraint came from the fact that
LV can shift the standard QED threshold for annihilation
of multi-TeV $\g$-rays from nearby blazars such as Mkn
501 with the ambient infrared extragalactic
photons~\cite{SG,Long,KonMaj,Giovanni,Comment,S03}.  LV depresses the
rate of absorption of one photon helicity and increases
it for the other.  Although the polarization of the $\g$-rays is not measured,
the possibility that one of the polarizations is
essentially unabsorbed
appears to
be ruled out by the observations
which show the predicted attenuation\cite{S03}.

{\em Electron and positron dispersion}.---
The Dirac equation in the Lorentz violating
EFT including the dimension five operators
can be written~\cite{MP} as
\beq
\left[i\sla{\partial} - m +
(\eta_1 \sla{u} + \eta_2 \sla{u} \g^5)
(u\cdot\partial)^2/M\right] \psi = 0,
\label{Dirac}
\eeq
where $u^\a$ is the unit timelike 4-vector that
specifies the preferred frame.  
If we choose coordinates aligned with $u^\a$,
so that $u^\a=\d^\a_0$, an electron or positron
mode
of energy
$E$ and momentum $p$ in the $x^3$ direction
contributes to the field operator via
$\exp(\mp i(Ex^0 - px^3))\Upsilon$, where
the upper sign here and below
is for an electron and the lower for a
positron, and $\Upsilon$ is the spinor.
Inserting this in the deformed Dirac equation
(\ref{Dirac}) yields
 \beq \left[\pm E\g^0 \mp p\g^3 - m
  - E^2(\eta_1\g^0 + \eta_2 \g^0\g^5)/M\right]\Upsilon =0.
\label{DiracEp}
\eeq
The helicity operator
acting on $\Upsilon$
is $\pm(p_i/|p|)\Sigma^i$~\cite{Peskin}, where
$\Sigma^i=\g^5\g^0\g^i$.
This is hermitian and
commutes with $\g^0$ times the operator in (\ref{DiracEp}),
which is also hermitian. Hence helicity remains
a good quantum number in the presence of this Lorentz violation.
Assuming without loss of
generality that $p>0$, a
spinor for
helicity $h$
therefore satisfies $\g^5\g^0\g^3 \Upsilon = \pm h
\Upsilon$, or equivalently $\g^0\g^5\Upsilon = \pm
h\g^3\Upsilon$.  For helicity eigenstates therefore
\beq
\left[(\pm E- \eta_1E^2/M)\g^0 -(\pm p\pm h\eta_2E^2/M)\g^3
  - m\right]\Upsilon =0.
\eeq
This has the form of the standard Dirac equation, with
$E$ replaced by $\tilde{E}=\pm E- \eta_1E^2/M$ and $p$
replaced by $\tilde{p}=\pm (p+h\eta_2E^2/M)$. Hence the
dispersion relation is given by $\tilde{E}^2 =
\tilde{p}_3^2 + m^2$.  For $m\ll p\ll M$ this yields
\beq
E^2 = p^2 +m^2 + 2(\pm \eta_1 +h\eta_2)E^3/M.
\eeq
With the definitions $\eta_R=2(\eta_1+\eta_2)$ and
$\eta_L=2(\eta_1-\eta_2)$, the parameters in the dispersion
relations for positive and negative helicity states
respectively are thus $\eta_R$ and $\eta_L$ for
electrons, and $-\eta_L$ and $-\eta_R$ for positrons.

{\em Possible new constraints from helicity decay}.--- If $\eta_R$
and $ \eta_L$ are unequal, say $\eta_R>\eta_L$, then a positive
helicity electron can decay into a negative helicity electron and
a photon, even when the LV parameters do not permit the vacuum
\v{C}erenkov effect. In this process, the large $R$ or small
($O(m/E)$) $L$ component 
of a positive helicity electron
transitions to the small $R$ or large $L$ component 
of a negative helicity electron 
respectively.  Such ``helicity decay" has no
threshold energy, so whether this process can be used to set
constraints on $\eta_{R,L}$ is solely a matter of the decay
rate.  It can be shown (assuming $|\xi|\lesssim10^{-3}$)
that for electrons of energy less than the
transition energy $(m^2M/(\eta_R-\eta_L))^{1/3}$, the lifetime of
an electron susceptible to helicity decay is greater than $4 \pi
M/(\eta_R-\eta_L)e^2 m^2$.  At the limit of the best current bound
$|\eta_L-\eta_R|<4$, the transition energy is approximately 10 TeV
and the lifetime for electrons below this energy is greater than
$10^4$ seconds.  This is long enough to preclude any terrestrial
experiments from seeing the effect.
The lifetime above the transition energy is instead
bounded below by
$E/e^2 m^2$, which is $10^{-11}$ seconds for energies just above
10 TeV. The lifetime might therefore be short enough to
provide new constraints.

Such a constraint might come from the Crab Nebula.
Suppose that $\eta_L$ is below the synchrotron constraint
(i.e. $\eta_L<-7\times10^{-8}$), so that $\eta_R$ must satisfy
both the synchrotron and \v{C}erenkov constraints
as explained above. Then
positive helicity electrons must have an energy of at
least 50 TeV to produce the observed synchrotron radiation.
These must not decay to negative helicity
electrons
(since those are unable to produce the synchrotron emission),
which would require that the transition energy
be greater than 50 TeV if the decay rate is fast enough.
This would yield the constraint
$\eta_R-\eta_L<10^{-2}$.

{\em Combined constraints}.--- The combined constraints are shown
logarithmically in Figure~\ref{fig:all}.
The vast improvement in
the birefringence constraint overwhelms the new synchrotron
\v{C}erenkov constraint, while the latter improves the previous
birefringence constraint~\cite{GK} by $10^2$. The allowed region
is defined above and below by the birefringence bound
$(O(10^{-14}))$, on the left by the synchrotron bound
$(O(10^{-7}))$, and on the right by the IC \v{C}erenkov bound
$(O(10^{-2}))$. If the polarization of GRB021206 proves incorrect,
the allowed region will expand vertically to the synchrotron
\v{C}erenkov lines. The combined constraints severely limit first
order Planck suppressed LV, making any theory that predicts this
type of LV very unlikely. The most useful improvements at this
stage would be to strengthen the positive $\eta$ and
$|\eta_R-\eta_L|$ bounds.

\section*{Acknowledgements}
This work was supported in part by the NSF under grants
PHY-9800967 and PHY-0300710, 
and by the CNRS at the Institut d'Astrophysique de Paris.

{\em Note added in proof.}---If the charges producing 
the Crab nebula gamma rays
consist of positrons as well as electrons, our
earlier argument implies only that one of the four
values $\pm \eta_{R,L}$ satisfies the
combined synchrotron and \v{C}erenkov constraints.
We are investigating whether a more complete analysis 
of the effect on the synchrotron and IC spectra 
provides a stronger constraint.


\end{document}